\documentclass{elsart}

\usepackage{harvard}

 \usepackage{epsfig}

\def\url#1{{\ttfamily\def\/{/\discretionary{}{}{}}#1}}

\def\farcs{\hbox{$.\!\!^{\prime\prime}$}}
\def\arcsec{\hbox{$^{\prime\prime}$}}
\def\lesssim{\mathrel{\hbox{\rlap{\hbox{\lower4pt\hbox{$\sim$}}}\hbox{$<$}}}}

\begin{document}

\begin{frontmatter}
\title{HST NICMOS Imaging of $z \sim$ 2 -- 3 radio-quiet quasars} 


\author[JHU]{S.\ Ridgway\thanksref{sr}}
\author[JHU]{T. Heckman\thanksref{th}}
\author[Calzetti]{D. Calzetti\thanksref{dc}}
\author[Lehnert]{M. Lehnert\thanksref{ml}}

\thanks[sr]{E-mail: ridgway@pha.jhu.edu}
\thanks[th]{E-mail: heckman@pha.jhu.edu}
\thanks[dc]{E-mail: calzetti@stsci.edu}
\thanks[ml]{E-mail: mlehnert@qso.mpe-garching.mpg.de}

\address[JHU]{Bloomberg Center for Physics and Astronomy, 
Johns Hopkins University, Baltimore, MD 21218}
\address[Calzetti]{Space Telescope Science Institute, 3700 San Martin Dr.,
Baltimore, MD 21218}
\address[Lehnert]{MPE, Postfach 1603, D-85740 Garching, Germany}

\begin{abstract}
We will present the results of a NICMOS H-band imaging survey of a small
sample of $z \sim$2 -- 3 radio-quiet quasars. We have resolved extension
in at least 4 of 5 objects and find evidence for a wide range in the 
morphologies and magnitudes of these hosts. 
The host galaxy luminosities range from sub-L$_*$ to about 4 L$_*$,
with most of the hosts having luminosities about L$_*$.
These host galaxies have magnitudes and
sizes consistent with those of the Ly break galaxies
at similar redshifts and at similar rest wavelengths, but
are about a magnitude fainter than the comparable 6C radio galaxies.
One residual host component is not centered on the quasar
nucleus, and several have close companions (within $\sim$10 kpc), 
indications that these systems are possibly in some phase of a 
merger process.  

\end{abstract}

\end{frontmatter}

\section{Introduction}
\label{intro}
Radio-quiet quasars (RQQ) represent more than half the known AGN at
high redshift ($z \stackrel{>}{_{\sim}} 1$), yet 
information on the galaxies that host these 
active nuclei is currently extremely limited. At lower redshift, the luminous
RQQ are found primarily in giant ellipticals with luminosities of
several times a $L_*$ galaxy at $z = 0$ (e.g. McLure et al. 1999).
In addition, the 
luminosity of the host seems to correlate roughly with that of the nucleus 
(McLeod \& Rieke 1995; McLure et al. 1999, McLeod,
Rieke, \& Storrie-Lombardi 1999).
Similarly, almost
every nearby bulge-dominated galaxy contains a supermassive
black hole candidate whose mass is roughly proportional to the mass of its
bulge (Magorrian et al. 1998; van der Marel 1999), implying a 
strong link between the formation and evolution
of galaxies and those of the quasars and their hosts. 
Semi-analytic hierarchical clustering models of galaxy formation
have been applied by Kauffman \& Haehnelt (1999) to this question,
and they have made some specific predictions about the
evolution of the nuclear magnitude--host relation for radio-quiet
quasars. To test such theories
we have made HST NICMOS observations of radio-quiet quasar hosts near the 
epoch of the peak quasar density ($z \sim$2 -- 3). We have chosen 
a sample of quasars whose nuclei
are faint enough to provide a good comparison sample
to those of the well-studied low-$z$ quasars. 
Throughout this paper we use  $H_0$ = 50 km s$^{-1}$ Mpc$^{-1}$ 
and $\Omega_0$  = 1. 

\section{Sample and Observations}

We selected 5 quasars from the faint quasar survey of Zitelli et al. (1992)
with 21.6$<$B$<$22.0 at $z \sim$ 2--3. The exact redshift range was constrained
to avoid emission lines falling in the NICMOS H filter passbands; this
resulted in a sample of 3 objects at $z \sim$1.8 and 2 objects at $z \sim$2.7.
Their nuclear $\rm M_{\rm B}$ are in the range
$-$22 to $-24$, making them comparable (and somewhat fainter) in absolute 
magnitude to many
low-$z$ quasar samples (e.g. Bahcall et al. 1997, McClure et al. 1999, McLeod
et al. 1999)
The observations were made using the NIC2 aperture of HST's NICMOS
camera, which has a field size of 19\farcs2 $\times$ 19\farcs2,
at a scale of 0\farcs075 pixel$^{-1}$.
To achieve emission-line-free imaging, the $z \sim 1.8$ objects were imaged
in the F165M filter, resulting in a restframe wavelength of $\sim$V,
and the $z \sim 2.7$ objects in the F160W filter, corresponding to
rest-frame $\sim$B.
We chose a nearby star for each of the 5 quasars, and observed this
star in the same visit as the quasar using the identical dither pattern
in order to characterize the point spread function (PSF).
The dither pattern included half-pixel offsets to improve resolution
by adequately sampling the HST PSF at these wavelengths.

We observed each of the 5 quasars (and its corresponding PSF star) with two
visits separated by several
months, resulting in observations of each field at significantly
different position angles on the sky relative to the PSF pattern
This allows an independent check
on the reality of any residual emission seen.
The final FWHMs achieved were 0\farcs14 -- 0\farcs16. 
Observing times per frame were on the order of 1500 s per quasar.
Total exposures for the quasars were
on the order of 6000 s per visit. However, due to variable
SAA CR persistence problems, sky noise levels in the final images
differ.  

\section{Data Reduction and Analysis}

We recalibrated the raw NICMOS data using a modified version of
the standard pipeline process.
We have combined each of the two visits to each object separately,
by using a simple method 
that determines the locations of the bad pixels in the initial 
frames and creates bad pixel masks using CRREJ, resamples the corrected frames 
to double the linear dimensions, and combines these using standard rejection.

We achieved the cleanest subtraction with a PSF star observed during the
same visit as the quasar.
We have made a simple, direct PSF subtraction through an iterative method that
varies the centering and scaling of the PSF versus the quasar, subtracts
the two images, and measures the chi-square of the residuals. 
Finding an unambiguous relative centering is simple; determining the
best scaling is more subjective. Here, we require flatness of
the residual across 
some central region (generally within a radius of $\sim$0\farcs15 ).
This will still likely result in an oversubtraction, depending
on how peaked the real host galaxy is within this inner radius.
In these images, the PSF residuals seem 
to dominate within a radius of $\sim$0\farcs1.
Within this region, we 
do not expect to recover morphological information; we can however find
real excess flux. 

\section{Results}
We have clearly resolved excess flux around 4 out of the 5 quasars relative
to the PSF stars, and also relative to a star with an apparent magnitude
similar to those of the quasars found within the field of MZZ 9592. 
This field star provides a good check that our results are not the product of
some difference in the way we have observed and reduced the PSF stars 
versus the quasars. We have thus also treated this star as if it were a 
quasar and applied the same PSF subtraction techniques to it; we find no
significant flux in the residual. 
In Figure 1, we show the results of these analyses for both visits 
of MZZ 9592, separately. We have rotated the results of
the second visit to match the orientation of the first visit. 
In Figure 2, we give the PSF-subtracted combined results
for 4 of the quasars, Gaussian-smoothed. This shows the range from the 
brightest to faintest of the residual host galaxies. 

To provide the simplest basis for cross-comparison between samples,
we have first calculated simple aperture magnitudes. We have 
used an aperture of
2\arcsec\ (corresponding to $\sim$16 kpc) to include most of the flux 
expected from a host galaxy, while excluding most nearby discrete companions.
To estimate errors, we have also calculated the host magnitudes for 
the two visits separately.
In most cases the differences were $\lesssim$0.3 mag.

We give in Table 1 the results of our magnitude analysis; we find the 
detected hosts vary from $\sim$L$_*$ to 4 L$_*$, using $L_*^V$ and $L_*^B$
at $z = 0$ from the field galaxy luminosity function 
of Loveday et al. (1992). The faintest residual host 
(around MZZ 4935) we may barely detect at a flux $\sim$0.4L$_*$.
These fluxes will generally need a correction for flux lost from the 
PSF subtraction process which will vary depending on how compact the
intrinsic, underlying host is. In the next section we discuss simple models
to estimate the amount of this correction.  

\begin{table}
\caption{MZZ quasar nuclear and host properties}
{\small
\begin{tabular}{l c c c c c c}
Name & Redshift & B & $\lambda_0$ & $M_B$ & $M_{\rm host}$ & Host Luminosity\\
& & & & & (Uncorrected)\\
MZZ4935  & 1.876 & 21.9 & 5737 \AA & -22.1 & -19.8 (V) & $\lesssim$0.4 $L_*^V$\\
MZZ11408 & 1.735 & 22.0 & 6033 \AA & -22.3 & -21.7 (V) & 0.9 $L_*^V$\\
MZZ1558 &  1.829 & 21.6 & 5832 \AA & -24.2 & -21.5 (V) & 0.8 $L_*^V$\\
MZZ9592 &  2.710 & 21.9 & 4312 \AA & -24.2 & -22.4 (B) & 3.6 $L_*^B$\\
MZZ9744 &  2.735 & 21.9 & 4284 \AA & -23.8 & -21.6 (B) & 1.7 $L_*^B$\\
\end{tabular}
}
\end{table}

The morphologies of these hosts are quite compact, generally less than
$\sim$0\farcs5 ($\sim$4 kpc) effective radius.
We also find in 2 cases galaxies close to the quasar in projection ($\lesssim$
10 kpc), and in MZZ 9592 an off-center host residual.  
\begin{center}
\begin{picture}(350,400)
\put(-80,-170){\includegraphics{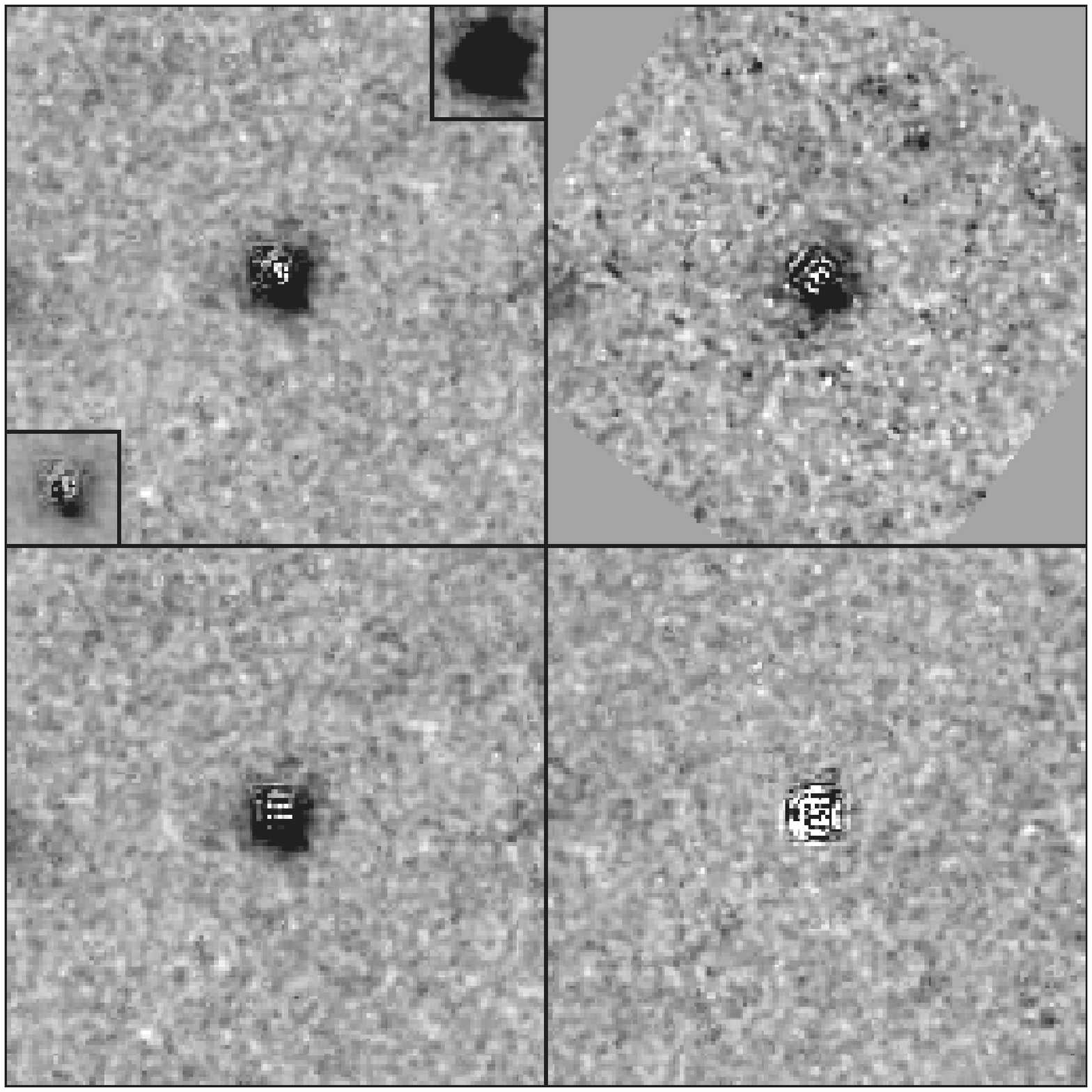}}
\put(-25,360){\large{\bf A.}}
\put(194,360){\large{\bf B.}}
\put(-25,130){\large{\bf C.}}
\put(194,130){\large{\bf D.}}
\end{picture}
\end{center}
\noindent
{\bf Figure 1.} MZZ 9592, PSF subtractions and tests.
{\it A.} Visit one, PSF-subtracted,
using the observed bright PSF star from the same visit.
The same image with a different scaling is shown in the lower left inset,
demonstrating that the residual host is not centered on the nucleus.
The upper right inset shows the unsubtracted quasar.
{\it B.} Visit two, PSF-subtracted, rotated to match the orientation of
visit one. Extra sky noise comes from unremoved low-level CR residuals.
{\it C.} Visit one, PSF-subtracted with the star that falls within the
field, demonstrating that the residual is not an artifact
of the mismatch between the observational
strategies applied to the quasar fields and the bright PSF stars.
{\it D.} The field star minus the PSF star: no residual flux.

\begin{center}
\begin{picture}(350,500)
\put(-80,-100){\includegraphics{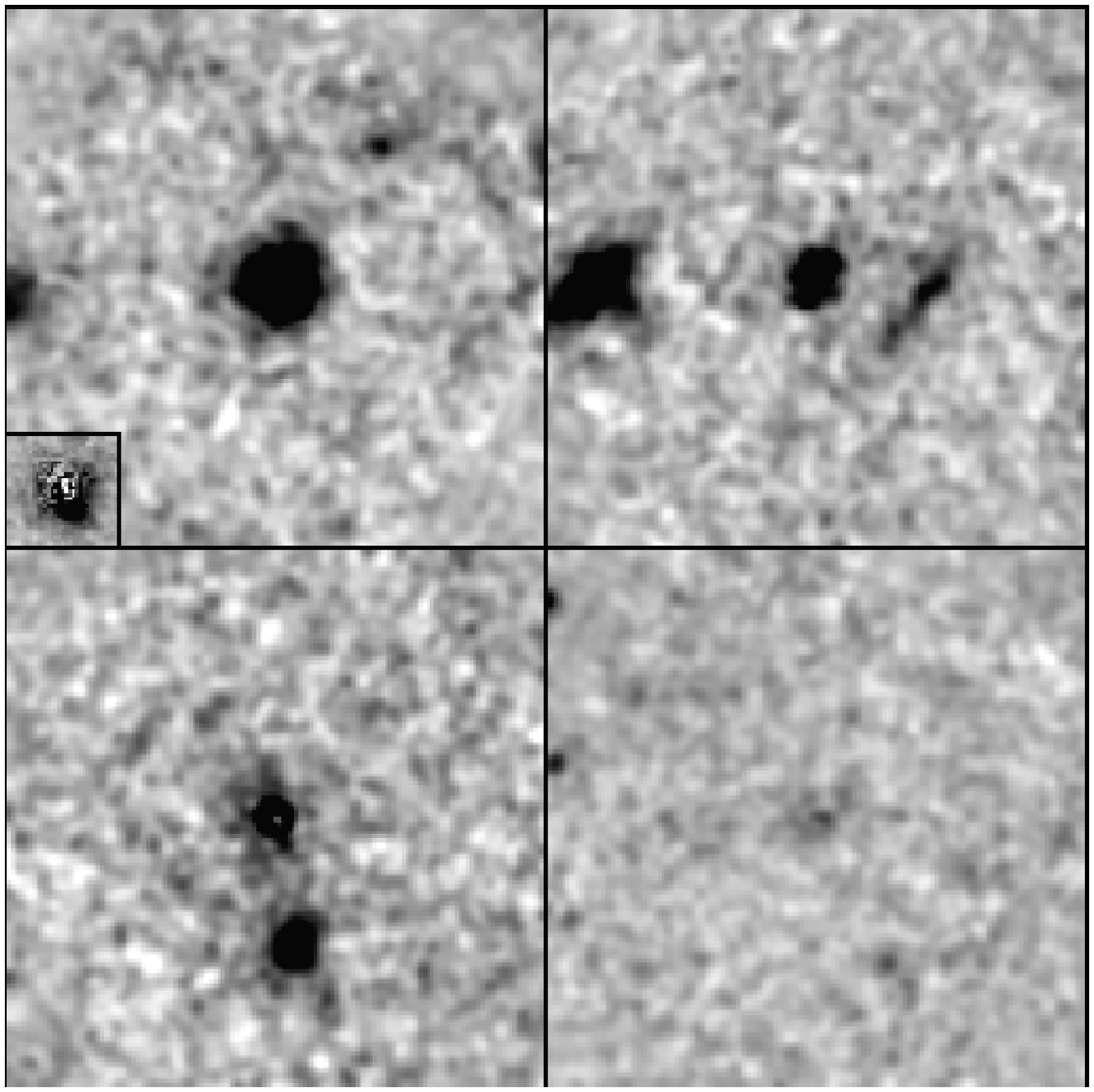}}
\put(-25,430){\large{\bf A.}}
\put(194,430){\large{\bf B.}}
\put(-25,200){\large{\bf C.}}
\put(194,200){\large{\bf D.}}
\end{picture}
\end{center}
 
\noindent
{\bf Figure 2.} PSF-subtracted MZZ quasar hosts, Gaussian-smoothed
with a kernel of 0\farcs06. Each panel is 5\farcs7 square
(or roughly 45 kpc), N up, E left.
{\it A.} MZZ 9592, $z \sim $ 2.7,
with an inset of the central region (unsmoothed).
There is an off-center residual host component.
{\it B.} MZZ 9744, $z \sim$ 1.8
{\it C.} MZZ 1558, $ z \sim$ 2.7 {\it D.} MZZ 4935, $z \sim $1.8,
host not detected. Note the apparent close ($\sim$ 10 kpc) companion
galaxies in panels {\it B} and {\it C}.

\subsection{Comparison to Lyman break galaxies and radio galaxies }

To compare the results of our study with studies of Lyman break galaxies
and radio galaxies at similar redshifts, we have made some simple simulations
of the effect the presence of the quasar nucleus and
the PSF-subtraction process have on the 
derived magnitudes and morphologies. We have used for the first set of these
simulations HST NICMOS observations
of a number of spectroscopically identified 
$z \sim$ 2 -- 3 Lyman break galaxies from the HDF North 
(Dickinson et al. 1999), which were observed in the F160W filter to a
much greater depth than our observations. For the observed galaxies
that were at the same redshift as our quasars, 
we first resampled these imaging data to match our final pixel scale, 
added ``quasar nuclei'' of varying magnitudes at 
the center of the galaxy by scaling and adding the MZZ9592 field star, then
added Poissonian noise to the appropriate level. We then used one of
the observed PSF stars to run the same automatic PSF-subtraction
process that we used on the quasars. 

We have found that the subtraction process in most cases on these
Lyman break galaxies resulted in a loss of flux of $\sim$0.3 magnitude, giving
us an indication of the amount to correct our derived hosts.
Even after such a correction, most of our quasar hosts have 
relatively moderate total magnitudes of $\sim$L$_*$, and these magnitudes 
and compact sizes are
basically consistent with the magnitudes and compactness
of star-forming galaxies at similar epochs (Dickinson et al. 1999).

To enable a comparison with previous ground-based infrared imaging
work on radio-loud galaxies, we adjust our fluxes to match those
of Eales et al. 1997, who studied 6C radio galaxies up to $z \sim$3. 
Extrapolating our H band fluxes to the K band gives K band magnitudes of
22.0, 20.3, and 20.5 for the $z \sim $1.8 quasar hosts, and 20.1 and
20.8 for the $z \sim$2.7 quasar hosts (after 
correction to the larger aperture used by Eales et al.).
They found that $z \sim 2$ 6C galaxies have a median 
$K$ magnitude of $\sim$18.5, while our 3 quasar hosts are at least a 
magnitude fainter than this, even allowing for a half magnitude
correction to our fluxes lost in the subtraction process.
Our $z \sim$2.7 quasar hosts are also 
a magnitude fainter at K than the 6C objects. 

\section{Discussion}

As the quasars in our sample have nuclear magnitudes that are comparable
to those of the well-studied
samples of low redshift quasars, we can make a direct comparison, and find
that these $z \sim 2$ -- 3 RQQ hosts are of similar or fainter M$_{V}$
as those of the sample of Bahcall et al. (1997) and McLure et al. (1997),
for example. 
However, if the host galaxies of luminous radio-quiet quasars
evolve passively into
giant ellipticals today, then at high redshift they should have had similar 
host magnitudes to the radio galaxies, rather than the moderately faint
hosts we have found.
Our results are consistent with other recent results on some brighter, 
lensed quasars; 
Rix et al. 1999 find that their sample of lensed $z \sim 2$ radio-quiet quasars
(de-magnified $M_B$ $\sim$ 24 -- 28)
also had hosts with comparably faint magnitudes.
Though inconsistent with passive evolution, our finding L$_*$ hosts
at $z \sim 2$ -- 3 agrees fairly well with the 
bottom-up hierarchical galaxy formation models of Kauffman \& Haehnelt (1999);
they 
predict median host luminosities that are somewhat below
present day L$_*$ for quasars at $z=2$ (and even fainter at $z=3$),
for quasars with the nuclear magnitudes of our sample. 
These hosts might therefore still be 
undergoing major mergers which would allow them to evolve into the present
day gEs associated with low-$z$ quasars; it is unclear in this 
interpretation, however, why the radio galaxies do not undergo 
similar mergers and evolution. Conversely, these high-$z$ quasar
hosts could be L$_*$ galaxies that will not significantly evolve
in luminosity. 


\end{document}